# NeRF-QA: Neural Radiance Fields Quality Assessment Database

Pedro Martin, António Rodrigues, João Ascenso, and Maria Paula Queluz

Instituto de Telecomunicações, Instituto Superior Técnico, University of Lisbon, 1049-001 Lisbon, Portugal
email: {pedro.martin, antonio.rodrigues, joao.ascenso, paula.queluz}@lx.it.pt

*Abstract*— **This short paper proposes a new database – NeRF-QA – containing 48 videos synthesized with seven NeRF based methods, along with their perceived quality scores, resulting from subjective assessment tests; for the videos selection, both real and synthetic, 360 degrees scenes were considered. This database will allow to evaluate the suitability, to NeRF based synthesized views, of existing objective quality metrics and also the development of new quality metrics, specific for this case.**

***Keywords—NeRF, video database, view synthesis, quality assessment***

## I. Introduction

Since 2020, Neural Radiance Fields (NeRF) methods have been proposed as a powerful technique for synthesizing novel views of a visual scene, from a set of input views [1,2]. Compared to traditional Depth Image-Based Rendering (DIBR) algorithms [3,4], NeRF methods achieve a significant quality increase in novel view synthesis [1,4,5], by representing 3D scenes as a radiance field approximated by one or more Multi-Layer Perceptron (MLP). The MLP's is trained using, as input, the spatial location and viewing direction coordinates of scene views, with the output being the color and opacity values corresponding to the input coordinates; thus, a continuous mapping between 3D points, along a given view direction, and their corresponding colors and opacity, is learned. By applying classic volume rendering to a trained NeRF, it is possible to project the output colors and opacities into a single image. With the development of NeRF methods, and seeking the rendering time reduction, neural-network-free methods have also been proposed, using direct optimizations of voxel grids features for scene representation [6,7].

The use of NeRF methods creates new types of artifacts on the synthesized views, notably the so-called floaters [8], besides flickering object edges which are also common to DIBR-based synthesis. In the NeRF related literature, five objective quality metrics have been typically used to evaluate the view synthesis results, namely PSNR, SSIM [9], MS-SSIM [10], LPIPS [11], and JOD [12]. However, the adequacy of these metrics to assess the fidelity and realism of the rendered views was not proven and, considering the floaters artifact, they may be not well suited for NeRF based synthesized images.

Recently, the authors of [13] have conducted a subjective quality evaluation study for NeRF based view synthesis considering only real scenes with front-facing views, i.e. the training images were acquired on a uniform grid, covering limited vertical and horizontal visual ranges. To the best of our knowledge, no subjective studies have assessed the quality of NeRF synthesized views for 360 degrees scenes – where the camera may rotate 360 degrees around a point, or move around an interest region – for real or synthetic scenes.

This paper proposes a new database – NeRF-QA – containing videos synthesized with seven NeRF based methods along with their perceived quality scores, resulting from a subjective test campaign; both real and synthetic 360 degrees scenes were considered. This database, available in [14], will allow to evaluate the suitability, to NeRF based synthesized views, of existing objective quality metrics and also the development of new quality metrics, specific for this case.

The remaining of this paper is organized as follows. In Section II, the framework used for views synthesis and their subjective quality assessment process, are described; a brief summary of the considered NeRF methods is also presented. Section III describes the video datasets used for the synthesis and the subjective test methodology. In Section IV the subjective test results are presented and analyzed, and main conclusions are drawn.

## II. NeRF Creation and View Synthesis & Assessment

In this section, the designed framework for the NeRF method training, NeRF based views synthesis and respective subjective quality assessment, is presented; the selected NeRF methods are also summarily described.

### *A. Framework*

Fig. 1 depicts the framework's pipeline designed for real scenes; the required changes for synthetic scenes are described at the end of this section. As can be figured out, the framework is composed by two main modules - train and test - corresponding, respectively, to the training of the NeRF method, and to the views' synthesis and assessment. The training views (or "training frames" in Fig. 1) are extracted from a video representing a 360 degrees scene. A section of the input video is also extracted to be used as reference ("reference video" in Fig. 1) on the subjective assessment tests; the considered section length is defined by the "test time interval". The reference video frames are further subject to pose estimation, enabling the video synthesis at the resulting coordinates with the already trained NeRF model. The procedures involved in the framework are described in the following:

- **Video preprocessing:** Consists of spatially downsampling and cropping the input video, using the FF-MPEG tools [15].
- **Frames selection:** Corresponds to the selection of frames from the preprocessed video according to: *i)* a set of $n$ uni formly spaced frames are selected from the input video, for NeRF training purposes and, *ii)* a set of adjacent frames

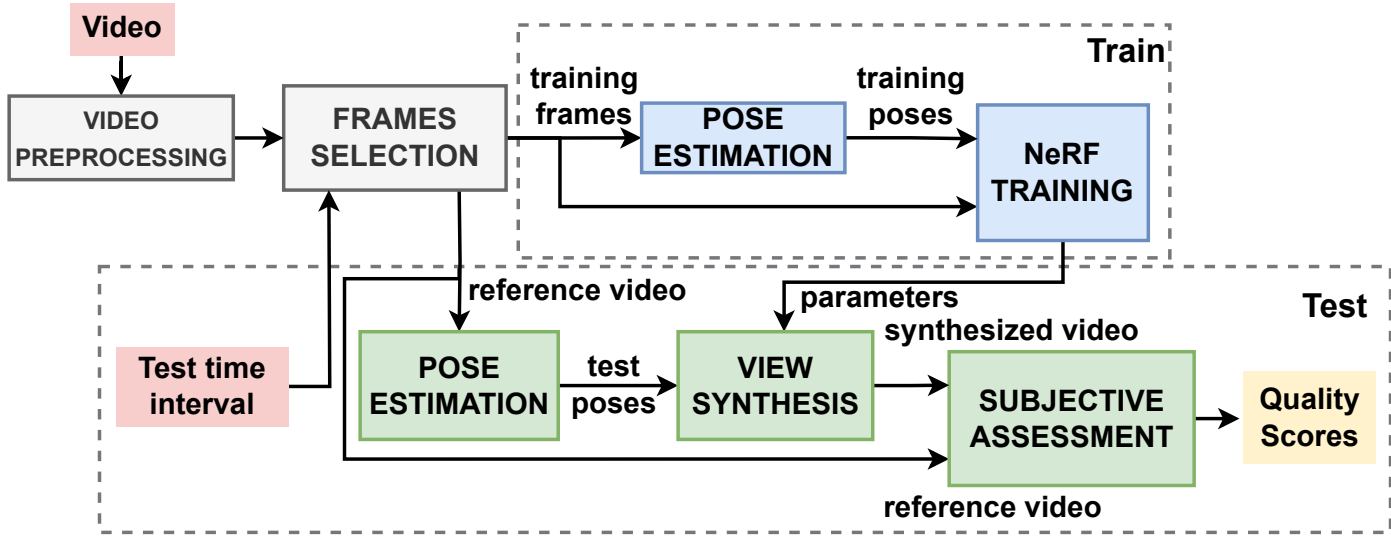

Fig. 1: Framework's pipeline for real scenes.

on a defined test time interval are extracted, to be used as reference video by the test module.

• **Pose estimation:** This process utilizes the COLMAP algorithm [7,8] to estimate the poses of the selected frames.

• **NeRF training:** Consists of an optimization procedure that aims to minimize a loss function between synthesized and reference images pixel values. A set of training images and the respective, pre-computed camera poses, are used for that purpose, enabling the method parameters tuning by minimizing the difference between the synthesized pixel values and the corresponding ones for the reference training images.

• **View synthesis:** This process involves querying to the trained NeRF method, for each pixel of the image being synthesized, the 5D coordinates (3D spatial location and 2D direction) of a set of sampling points along a camera ray (representing the scene intersection by that same pixel), about the respective color and opacity values. Finally, a classic volume rendering technique is applied to project the sampled points' colors and opacities into a novel image, corresponding to the requested viewpoint.

• **Subjective assessment:** The subjective quality assessment targets the perceptual quality evaluation of the synthesized views, by several human viewers. Several test methodologies are available, which are defined in international standards (mostly from ITU). This paper uses the Double Stimulus Continuous Quality Scale (DSCQS) method, defined in [18].

For synthetic scenes, the input scene is represented as a 3D model in Blender [19], which is used to render the training frames and the reference video. Thus, the video preprocessing, the frames selection, and the pose estimation are not applied, given that all these processes are suppressed by the use of Blender. An additional procedure corresponding to the subsampling of the training frames is added, seeking that the resulting synthesized views' quality also cover the lowest qualities range.

### B. Selected NeRF Methods

From the wide range of NeRF methods that have been proposed in literature, a subset was selected according to their popularity in the related scientific community:

• **DVGO [7]:** This method adopts a scene representation consisting of a density voxel grid for scene geometry, and a feature voxel grid, with a shallow network, for complex view-dependent appearance. It includes new techniques to increase the training convergence speed and the view synthesis quality, namely post-activation interpolation on voxel density, and several priors to increase the robustness of the optimization process. In this paper the last version of DVGO was used [20].

• **Instant-NGP [21]:** Compared to previously proposed NeRFs, this method achieves a significant speedup in training, by implementing neural graphics primitives with a small neural network, using a multiresolution hash table.

• **Mip-NeRF 360 [22]:** This method addresses the high challenging rendering of the so-called "unbounded" 360 degrees scenes (corresponding to arbitrarily large scenes, and where objects may exist at any distance from the camera). To overcome issues such as blurry renderings, the authors use a non-linear scene parameterization, online distillation, and a distortion-based regularizer.

• **NeRF++ [23]:** Aiming to improve the view synthesis fidelity for 360 degrees unbounded scenes, NeRF++ incorporates a novel hierarchical sampling scheme and a shape-prior regularizer.

• **Nerfacto [24]:** This method was developed by the Nerfstudio framework authors [24] by combining components from pre-existing methods – notably, the NeRF-- ray generation and sampling [25], the Mip-NeRF 360 scene contraction [22], the Instant-NGP hash encoding [21], and the NeRF-W appearance embedding [26] – seeking the best trade-off between speed and quality.

• **Plenoxels [6]:** This method proposes a view-dependent sparse voxel model that can achieve a fidelity level similar to the one obtained with the seminal NeRF, work [1] without relying on neural networks. Plenoxels achieves this by using a sparse voxel grid that is optimized using a reconstruction loss relative to the training images, along with a total variation regularizer.

• **TensoRF [27]:** This method models and reconstructs radiance fields of a scene using a 4D tensor that represents a voxel grid with per-voxel multi-channel features. Unlike NeRF methods that purely use Multi-layer Perceptrons (MLPs), TensoRF factorizes the 4D scene tensor into multiple compact low-rank tensor components using traditional CANDECOMP/PARAFAC decomposition.

The NeRF++, Nerfacto, and Mip-NeRF 360 methods were specifically designed and tested for real scenes datasets. The remaining methods did not have a specific type of scene target and were tested for both real and synthetic scenes.

## III. Subjective Assessment Study

This section describes the subjective test evaluation of NeRF based synthesized videos.

### A. Experimental Setup

The views synthesis and the subjective quality assessment of the resulting videos was conducted for eight videos taken from two popular datasets, namely *Tanks and Temples* [28] and *Realistic Synthetic 360º* [1], where the former resulted from real-world 360 degrees captures of large scenarios, and the latter was obtained with Blender [19]. The selected videos from *Tanks and Temples* were *M60*, *playground*, *train*, and *truck*. This dataset was used as processed in [23], where the number of training frames are equal to 277, 275, 258, and 226 frames, respectively, having spatial resolutions of 1077×546, 1008×548, 982×546, and 980×546 pixels. The scenes selected from *Realistic Synthetic 360º* were *drums*, *ficus*, *lego*, and *ship*, having all 100 training frames with spatial resolutions of 800×800 pixels. It is worthy to note that *drums* and *ship* videos are quite challenging, since they contain objects with non-Lambertian and specular reflection effects. For the subjective test purpose, the spatial resolutions of the real scenes were uniformized with a downsampling to 960×540 pixels, followed by a cropping to 928×522 pixels. For generating the reference videos, the test time interval was set to 10 s for real scenes, and to 6 s for synthetic scenes. In every case, the rendered camera poses do not coincide with the training poses. In particular, for the real scenes, the test time interval value was selected to an in between training frames interval. The NeRF methods selected for the real scenes were DVGO, Mip-NeRF 360, Nerfacto, NeRF++; the last three methods have been specifically designed for the view synthesis of real scenes. For the synthetic scenes, the selected synthesis methods were DVGO, Instant-NGP, Plenoxels, and TensoRF, all being methods targeting the view synthesis of synthetic scenes. The selected datasets have already been used in published works, enabling the validation of the herein generated synthesized videos, by comparison of the obtained objective quality metrics values (using PSNR and SSIM) with the values reported on those works. Lastly, the synthetic scenes were also synthesized for the case where a subsampling with a factor of 2 was applied to the training set, seeking synthesized video qualities covering the lowest qualities range.

### B. *Subjective Assessment Methodology*

The DSCQS [18] was selected as the subjective assessment method. In this case, the subjective test participant is presented with two side-by-side videos displayed on a monitor, one of which is the reference video and the other is its synthesized version, generated by one of the selected NeRF methods. The participant is not aware of which of the videos is the reference and the synthesized video is being displayed at each side and may watch two repetitions of the videos. The users are asked to evaluate each video using two continuous sliders (one for each displayed video) with a scale between 0 and 100, but where five qualitative scores – *Bad, Poor, Fair, Good,* and *Excellent* – are the visible scale marks to help the participants decision. A total of 48 pairs of stimulus (32 synthesized synthetic videos + 16 synthesized real videos, together with the respective original videos) were assessed, resulting in a test duration of approximately 30 minutes per participant. The test session objectives were briefly explained before, with a training session that preceded the main test. The subjective assessment was performed using an ASUS ProArt PA32UC-K 4K HDR monitor, with native resolution of 1920×1080 pixels. A total of 21 non-expert-viewers, aged between 16 and 61 years, have participated in the subjective assessment. After the test, the resulting scores were processed according to [18] to obtain Differential Mean Opinion Score (DMOS) values for each synthesized video. To check for outliers, the Z-scores were also computed, resulting in one participant scores' removal.

## IV. RESULTS AND CONCLUSIONS

Table I presents the performance of each NeRF based synthesis method, concerning the training time (Tr. time), rendering time (Re. time), and model size (#Param). These results were obtained on two NVIDIA GeForce RTX 4090 GPUs using the original methods' configurations published with the respective codes. Fig. 2 depicts the resulting DMOS values for the 48 synthesized videos contained in the NeRF-QA database (available in [14]), showing a full coverage of the visual quality that the higher the DMOS value, the lower the quality). Regarding the real scenes, the "excellent" quality levels were never achieved, revealing a less accurate view synthesis in this case. As discussed in [22,23], this is due to the fact that large and detailed scenes require an higher neural network capacity, and ambiguity on the synthesized content distance is easier to occur because of the inherent complexity of reconstructing a large scene from a small set of images.

TABLE I. NERF METHODS' PERFORMANCES.

| NeRF Method | Tr. time | Re. time | #Param |
|---|---|---|---|
| DVGO (w/ syn. scenes) | 2.0 mins | 0.36 mins | 4.1 M |
| DVGO (w/ real scenes) | 7.7 mins | 1.4 mins | 4.1 M |
| Instant-NGP | 4.9 mins | 0.4 mins | 12.6 M |
| Mip-NeRF 360 | ~ 30 hours | 23.7 mins | 9.9 M |
| Nerfacto | 7.2 mins | 1.2 mins | - |
| NeRF++ | ~ 19 mins | 22.6 mins | 2.4 M |
| Plenoxels | 3.5 mins | 0.6 mins | 10 M |
| TensoRF | 4.0 mins | 6.4 mins | 27 M |

Fig. 3 depicts the DMOS values of each synthesized video, discriminated by the used synthesis method. From its analysis it is possible to conclude that, for real scenes synthesis, the NeRF methods that took longer training and rendering times (cf. Table I) – which are also the MLP-based methods – achieved the best scores, evidencing a trade-off between performance and synthesis quality.

Finally, regarding the synthetic scenes, Fig. 3 reveals the existence of negative DMOS values, meaning that some synthesized videos had, on average, better subjective evaluations than the reference video. In contrast to real scenes results, for synthetic scenes there is a significantly higher variation of the resulting DMOS values, depending on the analyzed scene.

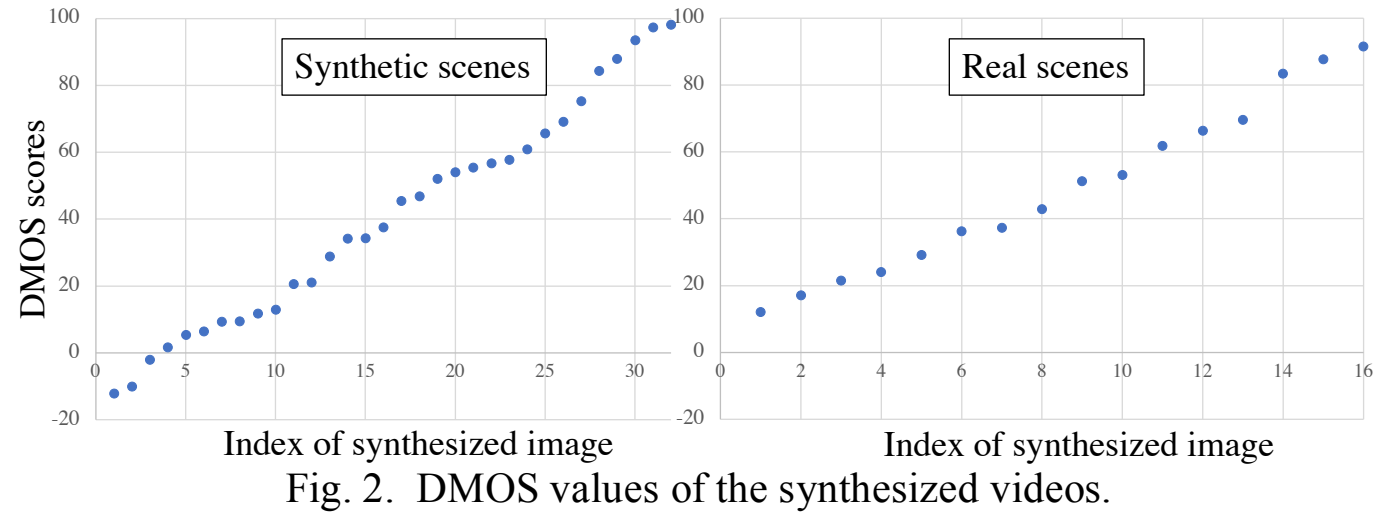

Fig. 2. DMOS values of the synthesized videos.

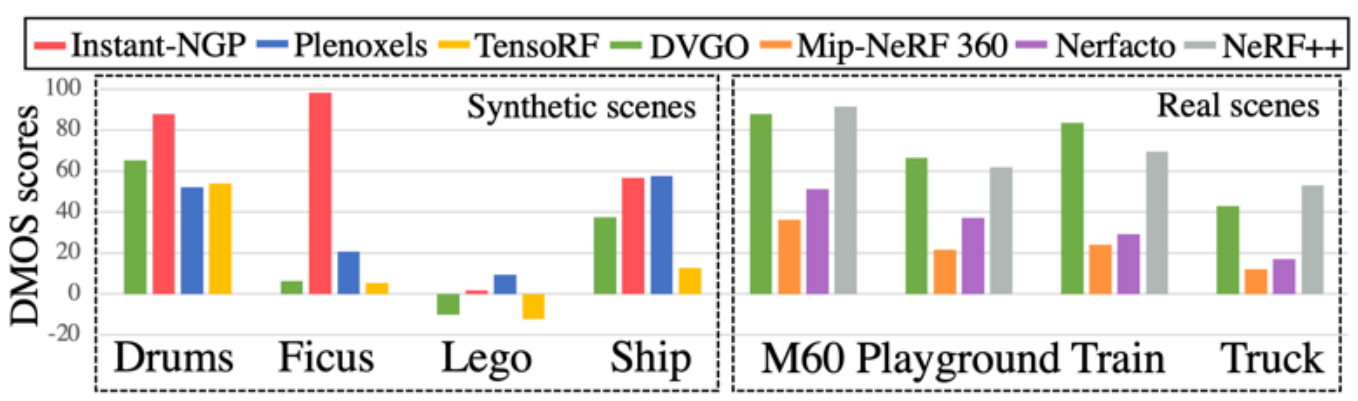

Fig. 3. DMOS values of each synthesized video, discriminated by the used synthesis method.